\documentclass[draftclsnofoot, onecolumn, 12pt]{IEEEtran}
\usepackage{cite}
\usepackage{graphicx}
\usepackage{psfrag}
\usepackage{subfigure}
\usepackage{amsmath}
\usepackage{amssymb}
\usepackage{epsfig}
\usepackage{color}
\usepackage{epsf}
\usepackage{algorithmic}
\usepackage{algorithm}
\usepackage{epsfig}
\usepackage{fancyhdr}
\usepackage{setspace}

\usepackage{fancyhdr}
\usepackage{amsfonts}
\usepackage{times}
\usepackage{latexsym}
\usepackage{dsfont}
\usepackage{amssymb}
\usepackage{amsmath}
\usepackage{cite}
\usepackage{verbatim}
\usepackage{subfigure}

\def\bb0{{\mathbb{0}}}

\def\bb{{\mathbf{b}}}

\def\bs{{\mathbf{s}}}

\def\bv{{\mathbf{v}}}

\def\b0{{\mathbf{0}}}

\def\bF{{\mathbf{F}}}

\def\bH{{\mathbf{H}}}

\def\bW{{\mathbf{W}}}

\def\sf0{{\mathsf{0}}}

\def\Nt{{N_\mathrm{T}}}
\def\Nr{{N_\mathrm{R}}}

\begin{document}

\title{The Practical Challenges of Interference Alignment\thanks{The authors at The University of Texas at Austin acknowledge the support of the Office of Naval Research (ONR) (Grant N000141010337), the DARPA IT-MANET program (Grant W911NF-07-1-0028), and the Army Research Labs (Grant W911NF-10-1-0420).}}

\author{\IEEEauthorblockN{Omar El Ayach, Steven W. Peters, and Robert W. Heath, Jr.\footnote{Omar El Ayach and Robert Heath are with The University of Texas at Austin, Austin, TX, USA. Steven Peters is with Kuma Signals LLC, Austin, Texas, USA. Robert Heath is the corresponding author (rheath@ece.utexas.edu).}\\}
}

\maketitle


\begin{abstract}

Interference alignment (IA) is a revolutionary wireless transmission strategy that reduces the impact of interference. The idea of interference alignment is to coordinate multiple transmitters so that their mutual interference aligns at the receivers, facilitating simple interference cancellation techniques. Since IA's inception, researchers have investigated its performance and proposed improvements, verifying IA's ability to achieve the maximum degrees of freedom (an approximation of sum capacity) in a variety of settings, developing algorithms for determining alignment solutions, and generalizing transmission strategies that relax the need for perfect alignment but yield better performance. This article provides an overview of the concept of interference alignment as well as an assessment of practical issues including performance in realistic propagation environments, the role of channel state information at the transmitter, and the practicality of interference alignment in large networks.

\end{abstract}


\section{Introduction} \label{sec:intro}

Interference is a major impairment to successful communication in commercial and military wireless systems. In cellular systems, interference is created when different base stations share the same carrier frequency due to frequency reuse. Inter-cell interference reduces data rates throughout the cells and causes outages at the cell edges. In local area networks, interference is created when different access points share the same channel. The medium access control protocol attempts to deal with interference by avoiding packet collisions (overlapping transmissions). This conservative approach leads to an under-utilization of system bandwidth. Similarly, neighboring nodes in a dense mobile ad hoc networks interfere if they share the same time and frequency resources. The medium access control protocol again limits the number of simultaneous conversations and consequently the system performance. Interference is thus a critical impairment in most wireless systems.

\begin{figure}[t!]
  \centering
	\includegraphics[width=6.4in]{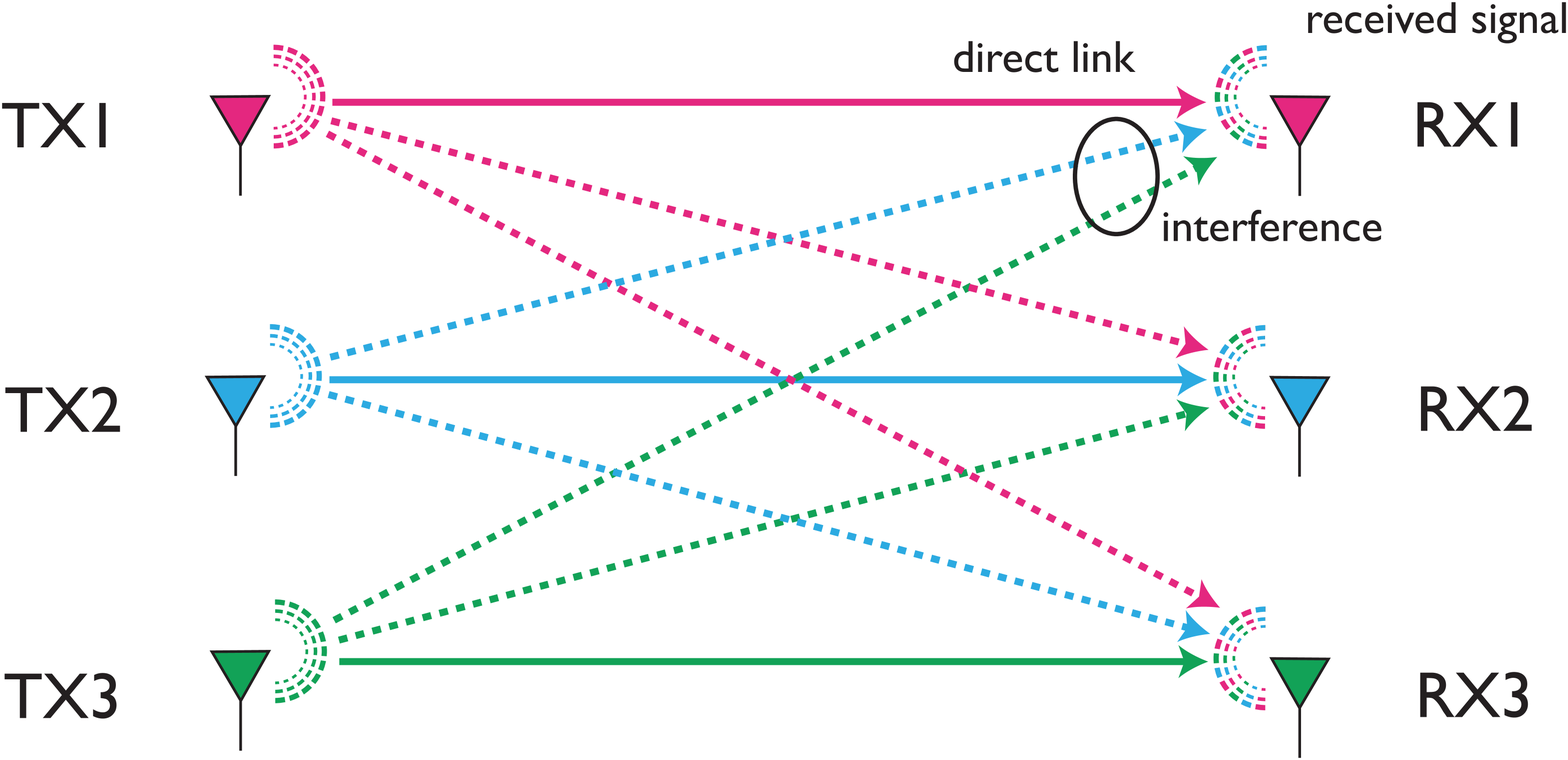}
	\caption{Illustration of a 3 transmit/receive pair interference channel. Each transmitter creates a signal that is interpreted as interference by its unintended receiver.}
	\label{fig:DARPA_IC}
\end{figure}

Communication in the presence of interference is often analyzed using an abstraction known as the interference channel. In the example interference channel of Fig. \ref{fig:DARPA_IC}, three different transmitters wish to communicate with three receivers. Each transmitter has a message only for its paired receiver. Assuming the transmitters share the same time and frequency resources, each transmission creates interference at the unintended receivers. There may be other sources of interference, not illustrated, such as jamming in military networks, or self-interference created from nonlinearities in the radio frequency components; those sources are not captured in the basic interference channel.

Traditional methods for dealing with interference often revolve around giving each user exclusive access to a fraction of the communication resources. In frequency division multiple access (FDMA), the system bandwidth is divided among the transmitters, e.g. in Fig. \ref{fig:DARPA_IC} each transmitter would be given a third of the total bandwidth. In time division multiple access (TDMA), transmitters take turns transmitting on a periodic set of transmission intervals.
Using a random access protocol, e.g. carrier sense multiple access, transmitters listen to see if the channel is available and then transmit if it is. Random access protocols are typically much less efficient than preassigned orthogonal access like FDMA or TDMA since the spectrum may not be fully utilized and collisions may occur. Regardless of the access protocol, the unifying concept remains avoiding interference by limiting the number of overlapping transmissions. If simultaneous transmissions are allowed, the resulting interference is often treated as noise and a loss in data rate ensues.

Recently a new concept for communication in interference channels, called interference alignment (IA), was proposed in \cite{Cadambe2008}. IA is a cooperative interference management strategy that exploits the availability of multiple signaling dimensions provided by multiple time slots, frequency blocks, or antennas. The transmitters jointly design their transmitted signals in the multi-dimensional space such that the interference observed at the receivers occupies only a portion of the full signaling space. An amazing result from \cite{Cadambe2008} is that alignment may allow the network's sum data rate to grow linearly, and without bound, with the network's size. This is in sharp contrast with orthogonal access strategies like FDMA or TDMA where sum rate is more or less constant since, regardless of network size, only one pair of users can communicate in a given time/frequency block.
Since the development of interference alignment, there has been work on a variety of topics to understand its theory and potential applications. A good summary of key results is given in \cite{jafar2011interference}.

The objective of this article is to review interference alignment with a focus on making it practical. First we describe IA in more detail, summarizing relevant practical issues that we believe are critical for its successful deployment. Then we review different techniques for computing alignment solutions, and more general interference management solutions. Because computing these solutions relies heavily on channel state information (CSI), we describe the two competing CSI acquisition techniques, reciprocity and feedback, and focus on their respective benefits and limitations. We highlight the fact that the dimensions needed for alignment, and the overhead of CSI acquisition, both rapidly increase with network size. This places a limit on the gains achievable via IA in large networks. For this reason we summarize work on partial connectivity and user clustering which leverages network topology information to reduce the requirements of alignment. We conclude with a discussion of the research challenges that remain in realizing interference alignment in practice. 

We note that the objective of this paper is to provide a high level introduction to the linear precoding type of interference alignment, which exploits channel state information for all users. Deeper discussions of other forms of interference alignment, including blind alignment, are found in \cite{jafar2011interference}.


\section{Linear Interference Alignment: Concept} \label{sec:IAconcept}

Interference alignment in its simplest form is a precoding technique for the interference channel. It is a transmission strategy that linearly encodes signals over multiple dimensions such as time slots, frequency blocks or antennas. By coding over multiple dimensions, transmissions are designed to consolidate, i.e. align, the interfering signals observed at each receiver into a low dimensional subspace. By doing so, interference alignment maximizes the number of non-interfering symbols that can be simultaneously communicated over the interference channel, otherwise known as the multiplexing gain. Interestingly, achieving the channel's maximum multiplexing gain, also known as degrees of freedom, implies that the sum rates provided by IA can approach sum capacity at high signal-to-noise ratio (SNR).

\begin{figure}[t!]
  \centering
	\includegraphics[width=6in]{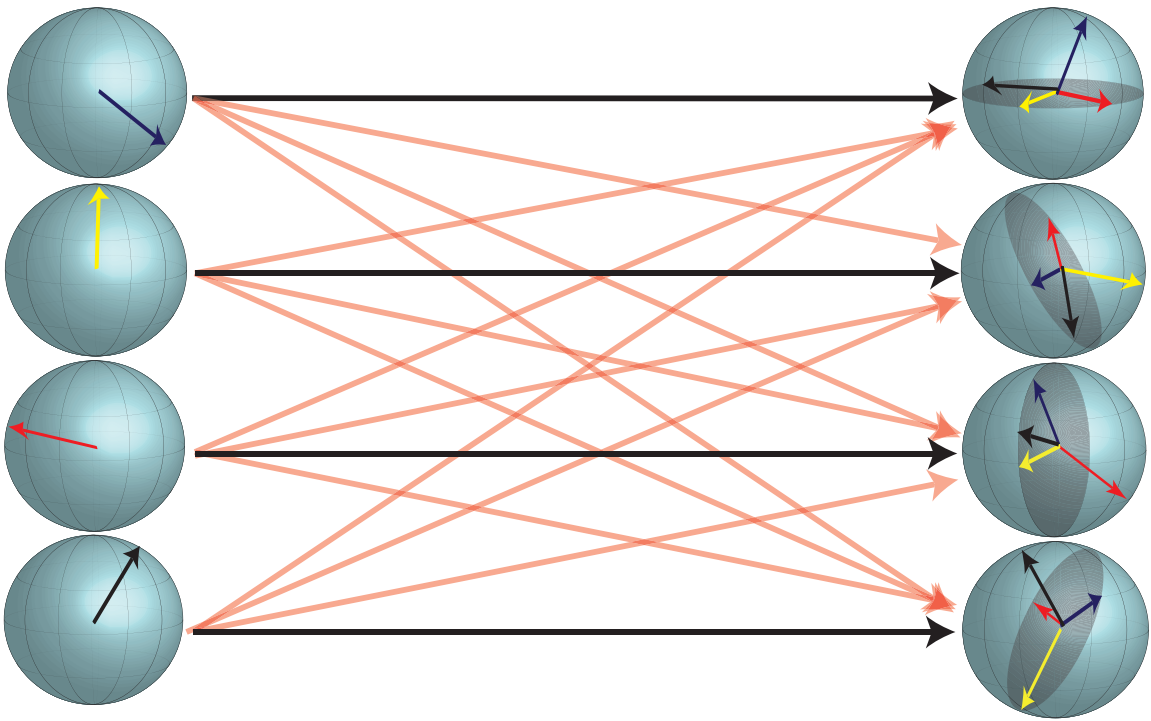}
	\caption{A diagram illustrating interference alignment. At each receiver, three interferers collapse to appear as two. This enables interference free decoding in a desired signal subspace.}
	\label{fig:IA_concept}
\end{figure}

To illustrate the IA concept, consider the four user system of Fig. \ref{fig:IA_concept} where real valued signals are coded over three dimensions and communicated over real valued channels. In this four user system, a receiver observes a total of three interference signals, each of which is represented as a vector in the real 3D space. Without careful structure, the three interference signals will occupy all three signal dimensions at the receiver; the signals do not fortuitously align. IA allows users to cooperatively precode their transmissions such that each \emph{three} interference signals are fully contained in a \emph{two} dimensional space. Alignment thus leaves one dimension in which users can decode their messages free from interference by projecting the received signal onto the subspace orthogonal to the interference subspace. 

The IA visualization in Fig. \ref{fig:IA_concept} translates into an equally intuitive mathematical representation. Consider a $K$ user interference channel in which each user $i$ transmits a set of $S_i$ information symbols encoded in the $S_i\times 1$ vector $\bs_i$. Symbols $\bs_i$ are precoded using a matrix $\bF_i$ and observed at each receiver $\ell$ after propagating over the matrix channel $\bH_{\ell,i}$; the dimensions of $\bF_i$ and $\bH_{\ell,i}$ will be made clear shortly. For such a system, the received signal at user $i$ is
\begin{equation}
\mathbf{y}_{i} = \mathbf{H}_{i,i}\mathbf{F}_{i}\mathbf{s}_{i} + \sum_{\ell \neq i} \mathbf{H}_{i,\ell}\mathbf{F}_{\ell}\mathbf{s}_{\ell} + \mathbf{v}_{i}, \nonumber
\label{eqn:narrowsig_model}
\end{equation} 
where the vector $\bv_i$ is the noise observed at receiver $i$. 
When only the time or frequency dimension are used for precoding, for example when coding over a bandwidth of $B$ subcarriers, $\bH_{i,\ell}$ is a $B\times B$ diagonal matrix; diagonal since subcarriers are orthogonal. In a MIMO system with $\Nt$ transmit antennas and $\Nr$ receive antennas, $\bH_{i,\ell}$ is $\Nr\times\Nt$ with no special structure.

Regardless of the dimension used to align interference, IA can be summarized as calculating a set of precoders $\bF_\ell$ such that any given user, even using a simple linear receiver $\bW_i$, can cancel the interference it observes from all other users, i.e., $\bW_i^*\bH_{i,\ell}\bF_\ell =\mathbf{0} \  \forall \ell \neq i,$ without nulling or destroying its desired signal  $\bW_i^*\bH_{i,i}\bF_i$. Early work on IA showed that a system's ability to find such IA precoders is directly related to the number of signal dimensions it can code over. Intuitively, the more time slots, frequency blocks, or antennas available for precoding, the more flexibility a system has in aligning interference. The problem of characterizing the number of coding dimensions needed to ensure the feasibility of IA has been studied extensively~\cite{Razaviyayn_Feasibility}. 




\section{Linear Interference Alignment: Challenges} \label{sec:IAchallenges}

Interference alignment relies on some assumptions which must be relaxed before it is adopted in practical wireless systems. In this section we briefly discuss some of the most pressing challenges facing the transition of IA from theory to practice.

\subsection{Dimensionality and Scattering} IA is achieved by coding interference over multiple dimensions. Intuitively, the more interfering signals that need to be aligned, the larger the number of dimensions needed to align them. 
When using the frequency domain for alignment, prior work has shown that the number of signaling dimensions needed to achieve good IA performance grows \emph{faster than exponentially} with the number of IA users. Properly aligning even as little as four users requires a potentially unreasonable number of subcarriers and a correspondingly large bandwidth~\cite{jafar2011interference}. This dimensionality requirement poses a major challenge for IA in practical systems.
The dimensionality requirement is relatively milder in the case of MIMO IA where more users can be supported as long as the number of antennas grows linearly with network size~\cite{Razaviyayn_Feasibility}. For this reason, IA seems most likely to be implemented in MIMO systems. We thus place a special focus on MIMO IA.
\subsection{Signal-to-Noise Ratio} 
IA is often degrees of freedom optimal, meaning that the sum rates it achieves approach the channel's sum capacity at very high SNR. At \emph{moderate SNR levels}, however, the sum rates resulting from IA may fall short of the theoretical maximum. As a result, IA may be of limited use to systems with moderate SNR unless IA algorithms are further improved. Examples of such algorithms are discussed in Section \ref{sec:IAalgos}.  
\subsection{Channel Estimation and Feedback} Channel state information (CSI), be it at the transmitter or receiver, is central to calculating IA precoders. As a result, sufficient resources must be allocated to pilot transmission, and in some cases to CSI feedback, to ensure the availability of accurate CSI. Since IA precoders must be recalculated when the channel changes appreciably, the overhead of CSI acquisition in high-mobility fast-fading systems can limit the gains of IA. For this reason, low overhead signaling strategies must be devised to properly tradeoff CSI quality with CSI acquisition overhead.
\subsection{Synchronization} IA via linear precoding is a transmission strategy for the \emph{coherent interference channel}. Thus, IA requires tight synchronization to remove any timing and carrier frequency offsets between cooperating nodes. In the absence of sufficient synchronization, additional interference terms are introduced to the signal model, rendering the IA solution ineffective. Synchronization strategies that leverage GPS satellite signals could help fulfill this requirement.
\subsection{Network Organization} Nodes cooperating via IA must not only synchronize, but also negotiate physical layer parameters, share CSI, and potentially self-organize into small alignment clusters if full network alignment proves to be too costly. In the absence of centralized control, distributed network protocols must be redesigned around this more complex and cooperative physical layer.

\section{Computing Interference Alignment Solutions} \label{sec:IAalgos}
There are some simple formulas for calculating IA precoders in some special cases~\cite{Cadambe2008}. To enable IA in general network settings, however, researchers have relied on developing iterative IA algorithms. Since then, the algorithmic focus has largely been on MIMO interference alignment.


The earliest method for finding MIMO IA precoders was the distributed solution presented in \cite{Gomadam2008}. The idea for the algorithm is that at each iteration, users try to minimize the extent to which their signal leaks into the other users' desired signal subspaces. After the algorithm converges, the IA condition $\bW_i^*\bH_{i,\ell}\bF_\ell=\mathbf{0}$ should ideally be satisfied and the desired signal spaces be free of interference. While the algorithm performs well at high SNR, it can be far from optimal in badly conditioned channels or at low SNR. The reason is that while this algorithm properly aligns interference, it is oblivious to what happens to the desired signal power during the process. 

The first attempt to improve low SNR performance is the Max-SINR algorithm of \cite{Gomadam2008}. Instead of minimizing interference power at each iteration, the improved algorithm accounts for the desired channel by iteratively maximizing the per-stream signal-to-interference-plus-noise ratio (SINR). By relaxing the need for perfect alignment, Max-SINR outperforms IA at low SNR and matches its performance at high SNR. Other algorithms have also relaxed the perfect alignment requirement and adopted more direct objectives such as maximizing network sum rate.
For example, \cite{luo_kien_algo} highlights the equivalence between maximizing sum rate and minimizing the signal's sum mean square error and uses the properties of the equivalent minimum mean square error (MMSE) problem to give a simple iterative algorithm in which precoders can be updated in closed form or via traditional convex optimization.

Suboptimal low SNR performance is admittedly not the only limitation of IA and many algorithms have been developed to address various shortcomings. Most early work on IA, for example, focused on the case where all interfering users are able to cooperate. The ability to cooperate, however, is fundamentally limited by constraints such as the number of antennas and cooperation overhead. As a result, large networks will inevitably have uncoordinated interference, or colored noise, which motivates the improved IA algorithms in~\cite{Peters2010}. Another limitation of precoding for the MIMO interference channel is that it often requires sharing entire channel matrices, potentially incurring a large overhead penalty as we will discuss in Section \ref{sec:CSI}. The work in \cite{IA_pricing} proposes a distributed transmission strategy inspired by game theory in which both precoders and transmit powers are iteratively adjusted to maximize sum rate by sharing \emph{scalar quantities} known as ``interference prices.'' The strategy in \cite{IA_pricing} thus replaces channel \emph{matrix} feedback with several iterations of \emph{scalar} feedback. Such a strategy could potentially reduce IA's overhead.

While editorial constraints have limited the discussion to the solutions of \cite{Cadambe2008,Gomadam2008,Peters2010, luo_kien_algo,IA_pricing}, a large number of noteworthy solutions have been developed to improve upon IA performance. The interested reader is encouraged to see the extensive list of IA-inspired algorithms and results in \cite{IA_bib}.


\section{Obtaining CSI in the Interference Channel} \label{sec:CSI}

Calculating IA precoders requires accurate knowledge of the interference generated by each transmitter. The premise is that if a transmitter knows the ``geometry'' of the interference it creates, it can conceivably shape it to mitigate its effect. Two methods of obtaining this knowledge are reciprocity and feedback. 

\subsection{Interference Alignment via Reciprocity} \label{sec:reciprocity}
In time division duplexed systems where transmissions on the forward and reverse links overlap in frequency and are minimally separated in time, propagation in both directions will be identical. In such systems, channels are said to be reciprocal. Reciprocity enables IA by allowing transmitters to infer the structure of the interference they \emph{cause} by observing the interference they \emph{receive}. 

Consider for example the IA strategy proposed in \cite{Gomadam2008} in which the precoders $\bF_i$ and the combiners $\bW_i$ are updated iteratively to reduce the power of uncanceled interference $\bW_{i}\bH_{i,\ell}\bF_\ell$ to zero. To do so, transmitters begin by sending pilot data using an initial set of precoders. Receivers then estimate their interference covariance matrix and construct combiners that select the receive subspaces with least interference. When the roles of transmitter and receiver switch, thanks to reciprocity, that same \emph{receive subspace} that carried the least interference becomes the \emph{transmit direction} which causes the least interference on the reverse link. Iterating this subspace selection on the forward and reverse links results in a set of precoders satisfying the IA conditions.

\begin{figure}[t!]
\centering
\subfigure[Reciprocity based IA]{
\includegraphics[width=3.1in]{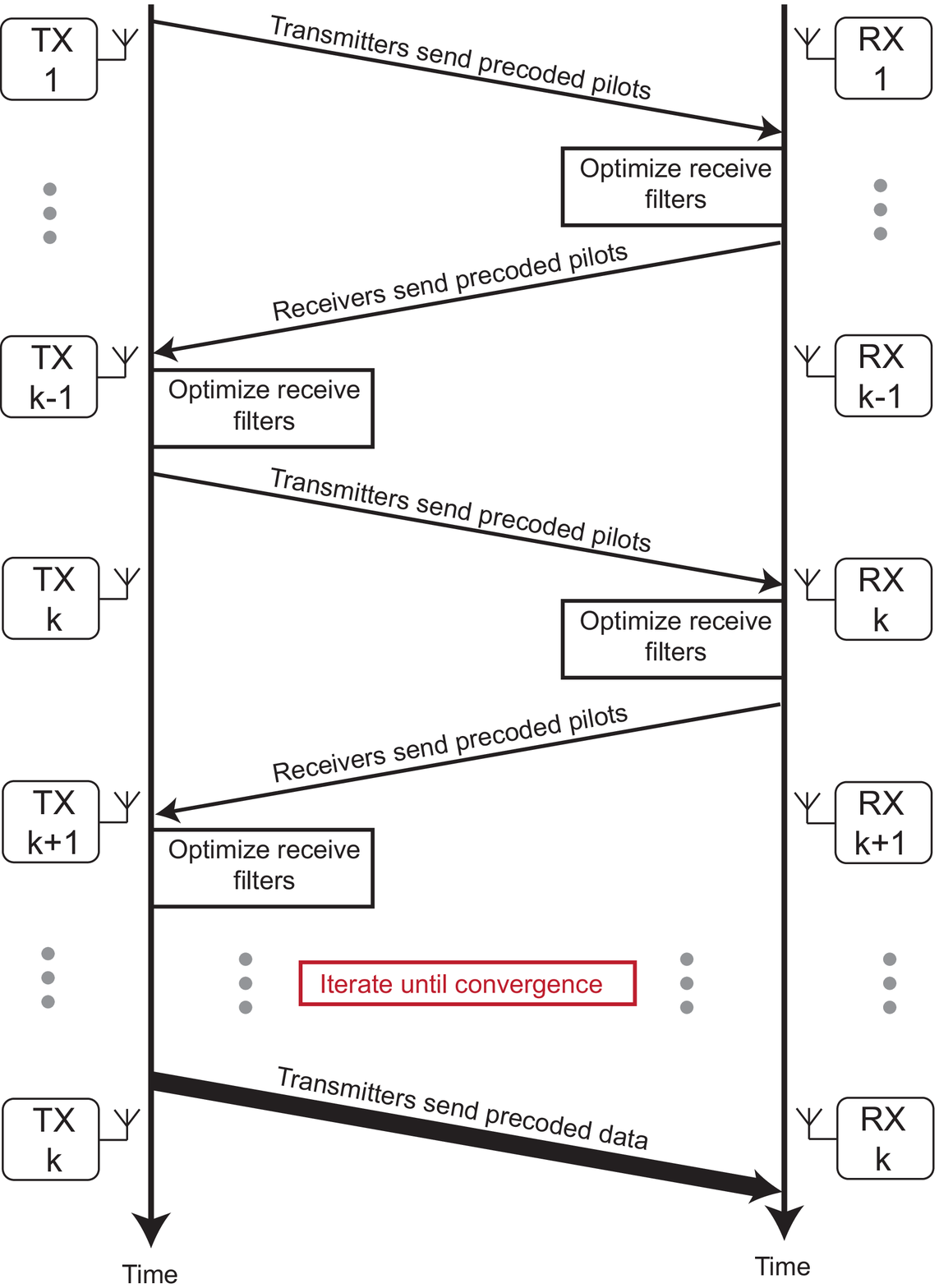}
\label{fig:reciprocity}
}
\subfigure[Feedback based IA]{
\includegraphics[width=3.1in]{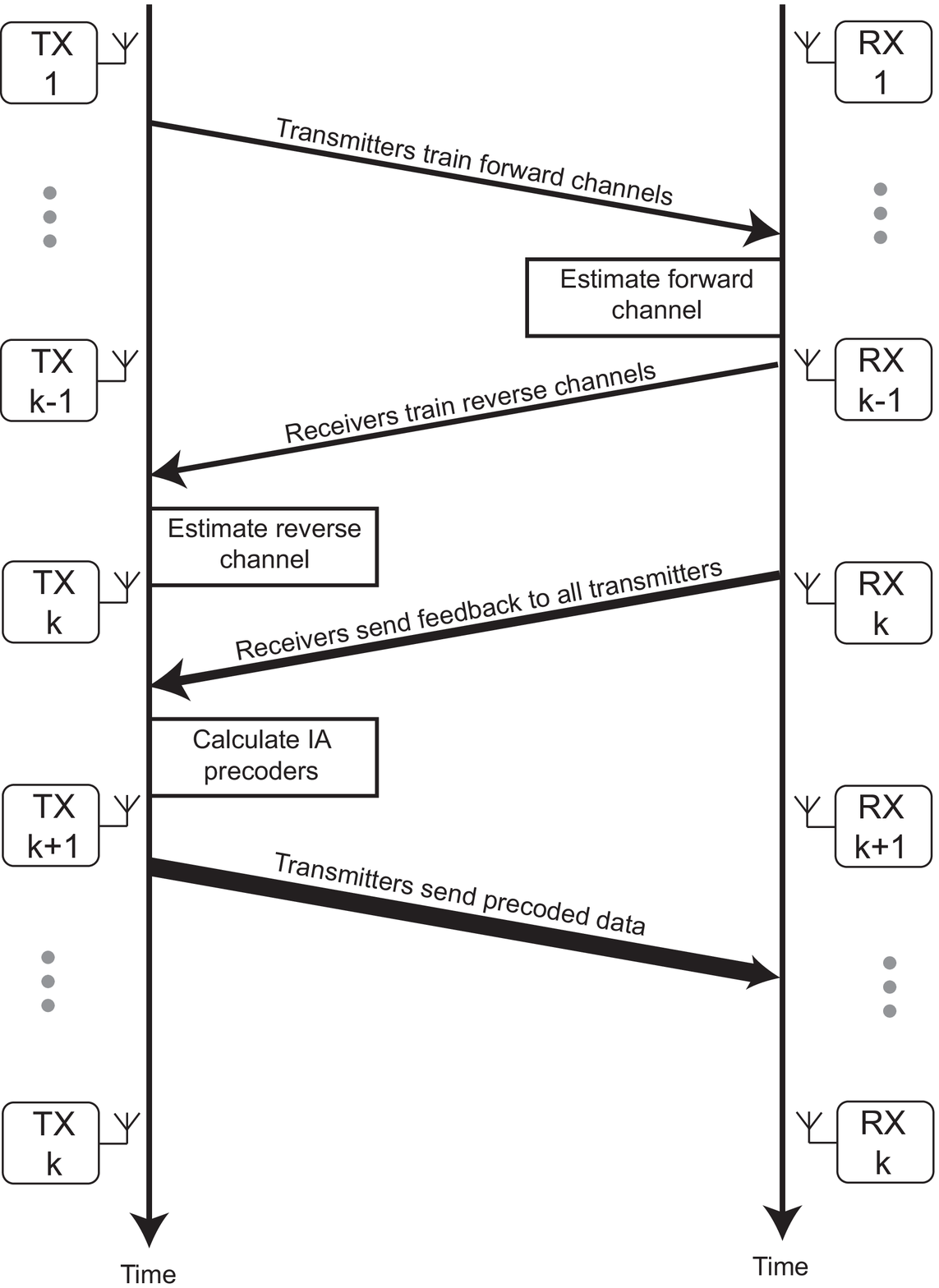}
\label{fig:feedback}
}
\label{fig:subfigureExample}
\caption{The two CSI acquisition paradigms considered in IA literature: (a) Reciprocity based strategies which infer the forward channel from reverse link pilots, and (b) feedback strategies which rely on explicitly communicating forward channel information to the transmitters.}
\end{figure}

While \cite{Gomadam2008} at each iteration chooses the subspace with least interference, a variety of more sophisticated objectives may be considered. For example \cite{Berry-BidirectionalIA} considers an update rule that minimizes the signal's mean square error, resulting in improved sum rate performance. Regardless of the subspace selection rule, the general framework for precoding with reciprocity is shown in Fig. \ref{fig:reciprocity} and proceeds as follows:
\begin{enumerate}
\item \emph{Forward link training}: Transmitters send precoded pilot symbols using a set of initial precoders. Receivers estimate forward channel parameters and compute combiners that optimize a predefined objective. 
\item \emph{Reverse link training}: Receivers send precoded pilot symbols using the combiners from step 1 as transmit precoders. Transmitters in turn optimize their combiners/precoders and initiate a second training phase with the updated precoders.
\item Communicating pairs iterate the previous steps until convergence.
\item \emph{Data transmission}: Payload data is then communicated.
\end{enumerate}

Relying on reciprocity for precoding over the interference channel has a number of potential drawbacks. First, iterating over the air incurs a non-negligible overhead due to the recurring pilot transmissions. While the results in \cite{Berry-BidirectionalIA} consider pilot overhead, more work is needed to settle the viability of reciprocity. Second, reciprocity may not suffice for all IA-based algorithms. 
For example, one of the algorithms in \cite{Peters2010} attempts to improve IA by considering sources of uncoordinated interference. Since the uncoordinated interference observed by the transmitters and receivers is not ``reciprocal'', reciprocity cannot be used. Third, reciprocity does not hold in frequency duplexed systems and ensuring reciprocity with time duplexing requires tightly calibrated RF devices.

\subsection{Interference Alignment with Feedback} \label{sec:feedback}

Several IA results have also considered systems with CSI feedback. A general feedback framework is shown in Figure \ref{fig:feedback}. In such a system, the transmitters first send training sequences with which receivers estimate the forward channels. Receivers then feedback information about the estimated forward channels, potentially after training the reverse link. After feedback, the transmitters have the information needed to calculate IA precoders. Feedback, however, invariably introduces distortion to the CSI at the transmitters and incurs a non-negligible overhead penalty. Therefore, the difficulty lies in designing low-overhead low-distortion feedback mechanisms for IA.

An established method to provide high quality feedback with low overhead is limited feedback, i.e., channel state quantization. Limited feedback was first considered in \cite{Thukral2009} for single antenna systems where alignment is achieved by coding over OFDM subcarriers. The feedback strategy in \cite{Thukral2009} leverages the fact that IA solutions remain unchanged if channels are scaled or rotated which allows efficient quantization via what is known as Grassmannian codebooks. Unfortunately, maintaining proper alignment requires that the accuracy of quantized CSI improve with SNR which in turn implies that the quantization codebook size must scale \emph{exponentially} with SNR. The complexity of quantized feedback, however, increases with codebook size and large Grassmannian codebooks are difficult to design and encode. Furthermore, this strategy relies on the CSI's Grassmannian structure for efficient quantization. As a result, it cannot be applied to systems where the CSI exhibits no special structure. This alienates a main case of interest, namely IA in multiple antenna systems where the CSI to be fed back is the set of channel matrices.

To support MIMO IA and overcome the problem of scaling complexity, IA with analog feedback was considered in~\cite{Ayach2010a}. Instead of quantizing the CSI, analog feedback directly transmits the channel coefficients as uncoded quadrature and amplitude modulated symbols. Since no quantization is performed, the only source of distortion is the thermal noise introduced during training and feedback. As a result, the CSI quality naturally increases with the SNR on the reverse link. This implies that IA's multiplexing gain is preserved as long as the forward and reverse link SNRs scale together. 

Feedback, however, poses a number of challenges. Not only does the required quality of CSI scale with SNR, but the required quantity also scales super-linearly with network size. If feedback is inefficient, IA's overhead could dwarf its promised theoretical gains. Second, IA relies on sharing CSI with interfering transmitters: multicell systems that share transmitter CSI over a backhaul may suffer from queuing delays that render CSI obsolete on arrival. While work on addressing these issues is underway, most research still considers systems with a limited number of users. In large scale networks, however, practical challenges are amplified many-fold.

\section{Interference Alignment in Large Scale Networks} \label{sec:largenets}

Consider applying IA, or any other interference channel precoding strategy, to a large scale network, such as a regional cellular network. In this setting, the need for signaling dimensions grows, the overhead of CSI acquisition explodes, and the task of synchronization becomes daunting. This grim outlook, however, is a byproduct of a naive over-generalization of the basic interference channel. In large scale networks not all interfering links are significant and thus not all of them should be treated equally. 

To examine the feasibility of IA in large networks, \cite{guillaud2011interference} considered a large network wherein each user receives interference from a finite subset of users, e.g., the first tier of interferers in a grid-like cellular system. Using this model, \cite{guillaud2011interference} showed that the number of antennas needed for network-wide alignment is a linear function of the size of the interfering subset rather than the size of the entire network. This implies that perfect alignment in an \emph{infinitely} large network is theoretically possible with a \emph{finite} number of antennas. This partially connected model, however, is ultimately a simplification of network interference. While some interferers may indeed be too weak to merit alignment, that threshold remains unclear. Moreover, neglecting interferers that fall below that threshold, as is implicitly done in the partially connected model, may not be optimal. Exploring the optimal transmission strategy in this gray area between full connectivity and true partial connectivity is the focus of \cite{Peters2010a}.

The work in \cite{Peters2010a} considers a wireless network and assigns a finite channel coherence time and different pathloss constants to each link. This setup allows \cite{Peters2010a} to gauge the gain from aligning a set of users versus the CSI acquistion overhead involved. On the one hand, it is noticed that the large coherence time in static channels dilutes the cost of CSI and enables alignment over the whole network. On the other hand, fast fading channels do not allow enough time to acquire the CSI needed for IA. In this latter case, systems perform better with lower overhead strategies such as TDMA. In between the two extremes, it is noticed that systems benefit most from a hybrid IA/TDMA strategy wherein smaller subsets of users cooperate via alignment, and TDMA is applied across groups. The objective then becomes finding an optimal user grouping, or network partitioning, for which \cite{Peters2010a} provides various algorithms based on geographic partitioning, approximate sum rate maximization, and sum rate maximization with fairness constraints. In all grouping solutions only long-term pathloss information is used which further reduces the overhead of IA by avoiding frequent regrouping.

The IA research community has made successful initial attempts at demonstrating that the reach of IA can extend well beyond the information theoretic interference channel through work such as \cite{guillaud2011interference, Peters2010} and many others \cite{IA_bib}. Perhaps the most crucial next step towards realizing IA gains is to rethink the upper layers above IA in the communication stack, e.g., the medium access layer. These protocols have traditionally been built around point-to-point physical layers and must be redesigned to enable more cooperative multi-user physical layers instead. Preliminary work on this front is already in progress through the prototyping work in \cite{lin2011random} which extends the earlier prototyping efforts by the same authors.

\section{A Practical Performance Evaluation} \label{sec:sims}
\begin{figure}[t!]
  \centering
	\includegraphics[width=4.4in]{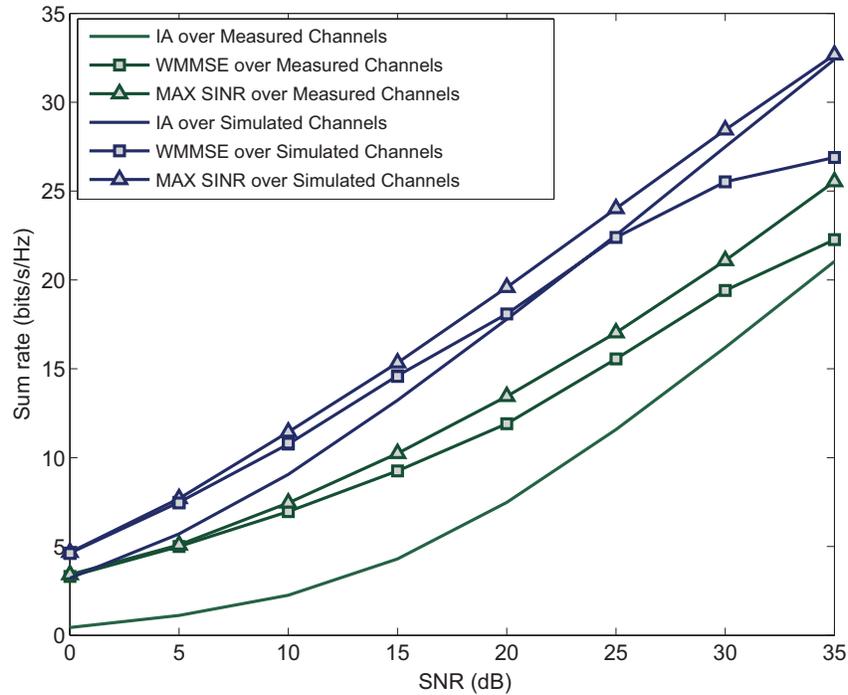}
	\caption{A performance comparison for three precoding solutions for the interference channel over both measured and simulated channels.}
	\label{fig:algo_comparison}
\end{figure}

In this section we provide numerical results on the performance of IA with a special focus on the concepts discussed throughout the article. Fig. \ref{fig:algo_comparison} compares the performance of the algorithms in Section \ref{sec:IAalgos} in a three user system with two antennas per node. For this assessment both simulated channels and measured MIMO-OFDM channels from the testbed in \cite{Ayach2010} are considered. Simulated channels are assumed to be Rayleigh fading, independent across all users and antennas. This constitutes a baseline of highly scattered channels, considered ideal for IA. To demonstrate the effect of limited scattering, we report measurements from \cite{Ayach2010} corresponding to a scenario in which antennas are spaced half a wavelength apart. 

The results in Fig. \ref{fig:algo_comparison} verify that IA, in this case, achieves a multiplexing gain of three as predicted in theory. Multiplexing gain can be calculated by examining the slope of the sum rate curves. The same results, however, also demonstrate IA's sub-optimality at low SNR where precoding algorithms such as the WMMSE algorithm in \cite{luo_kien_algo} or the MAX-SINR algorithm in \cite{Peters2010} consistently outperform. 
Finally, Fig. \ref{fig:algo_comparison} shows that the correlation present in measured channels can have significant effects on the performance of IA. Interestingly, in such measured channels, the algorithms in \cite{Peters2010, luo_kien_algo} provide large gains over IA even at a significant SNR of 35dB.

\begin{figure}[t!]
  \centering
	\includegraphics[width=4.4in]{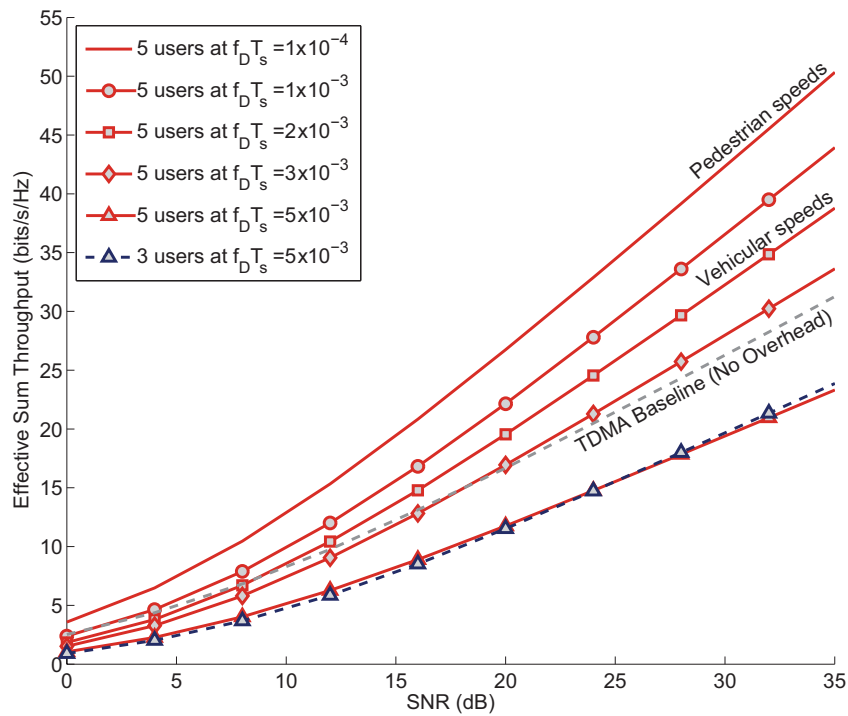}
	\caption{Effective throughput vs. SNR for a five user system at varying normalized Doppler spreads. Each transmitter is equipped with 3 antennas, and every user communicated via one spatial stream. The figure demonstrates how as Doppler increases, so does overhead and CSI error, resulting in a drop in effective sum rate.}
	\label{fig:IAoverhead}
\end{figure}

\begin{figure}[t!]
  \centering
	\includegraphics[width=4.5in]{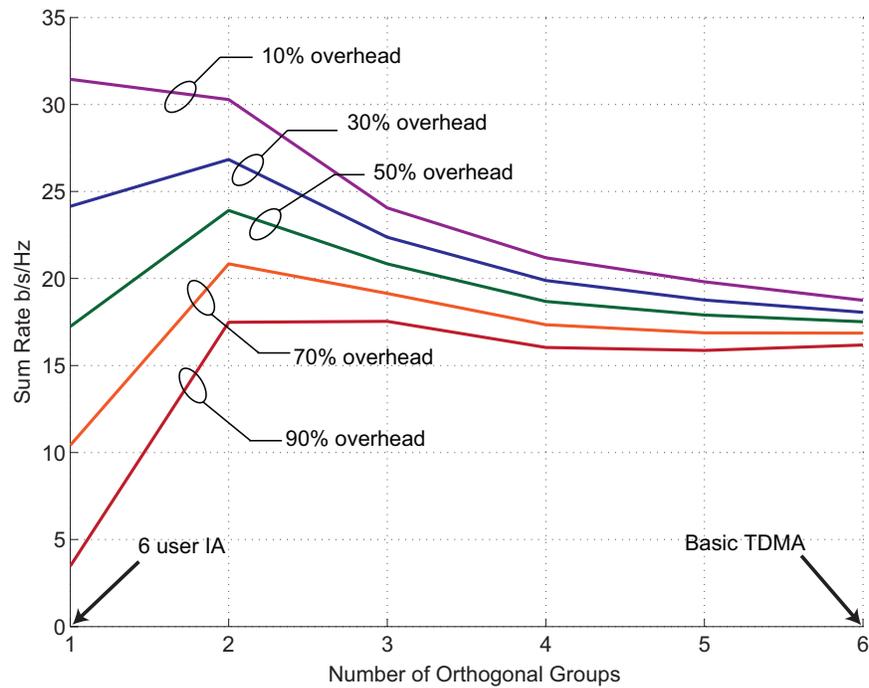}
	\caption{The effective sum rate achieved by full network IA, TDMA, or a partitioned network in which smaller groups coordinate via IA. As overhead increases, the plot indicates that this is often optimal to allow fewer users to communicate simultaneously. In highly }
	\label{fig:IAclusters}
\end{figure}

In Fig. \ref{fig:IAoverhead} we relax the assumption of perfect channel knowledge and consider a five user system that acquires CSI through training and analog feedback thus suffering from both CSI distortion and CSI acquisition overhead. The results indicate that, at pedestrian speeds, the slow channel variation enables acquiring accurate CSI at a low overhead cost which translates into substantial gains over non-cooperative transmission strategies such as TDMA. The same results indicate that in a high mobility setting, where CSI acquisition is costly, one may elect to either coordinate a smaller user group or altogether adopt a lower overhead strategy such as TDMA. In a large wireless network, grouping users based on overhead constitutes selecting from a continuum of network structures. For example, Fig. \ref{fig:IAclusters} demonstrates how even a simple six user system should be partitioned differently as the overhead cost of full IA varies. The throughput maximizing strategy transitions from full six user IA in static channels, to basic TDMA for fast fading channels, passing through a range of ``hybrid'' networks structures in which users cluster into small cooperating groups.


\section{Future Research Directions} \label{sec:futurework}
In this section we present some active areas of interest and general topics for further research on interference alignment.
\begin{enumerate}
\item \emph{Algorithms}: Algorithms remain a hot topic for IA research as a number of ``extra features'' can be incorporated or further improved such as complexity, low SNR performance, distributed-ness, robustness to CSI imperfections, etc.  
\item \emph{Feedback}: 
Temporal correlation can be leveraged to both improve the accuracy of CSI as well as reduce the overhead of feedback. Limited feedback strategies such as \cite{Thukral2009} can be generalized to better support MIMO IA. The development of limited feedback strategies can benefit from the lessons learned in the MIMO single user and broadcast channels where feedback can be significantly compressed by exploiting the mathematical properties of the quantized CSI. Finally, additional work is needed to better understand the effects of feedback overhead and feedback delay.
\item \emph{IA in Multihop Networks}: Preliminary work on multihop IA suggests that relays can greatly reduce the coding dimensions needed to achieve a network's degrees of freedom, and otherwise simplify the optimal transmission strategies~\cite{jafar2011interference}. The importance of practical precoding algorithms for the relay aided interference channel is further amplified by the standards community's growing interest in deploying relays in future wireless systems.
\item \emph{IA in Modern Cellular Networks}: 
To make IA a viable multi-cell cooperation technique, IA research must characterize the effect of cellular system complexities such as scheduling, resource allocation, and backhaul signaling delay. IA must also be re-evaluated using models that more accurately resemble modern cellular systems which could include heterogeneous infrastructure such as macrocells, picocells, small cells, relays, and distributed antennas.


\end{enumerate}


\section{Conclusion} \label{sec:conclusion}
Interference alignment is a transmission strategy that promises to improve throughput in wireless networks. This article reviewed the key concept of linear interference alignment, surveyed recent results on the topic, and focused on bringing the concept closer to implementation by discussing some of its main limitations. After the key hurdles in areas such as low-SNR performance, CSI acquisition overhead, synchronization and distributed network organization can be overcome, interference alignment will be ready for practical implementation. 




\bibliographystyle{IEEEtran}
\bibliography{IEEEabrv,IA_CommMag}

\newpage


\end{document}